\documentclass[traditabstract]{aa} 

\usepackage{txfonts} 
\usepackage{natbib} 
\usepackage{graphicx}

\begin{document}

\title{Lithium production in the merging of white dwarf stars}

\author{Richard~Longland\inst{1,2}\thanks{\email{richard.longland@upc.edu}}
  \and Pablo~Lor\'{e}n-Aguilar\inst{3,2} \and Jordi~Jos\'{e}\inst{1,2}
  \and Enrique~Garc\'{\i}a--Berro\inst{3,2} \and
  Leandro~G.~Althaus\inst{4}}
        
\institute{Departament de F\'{\i}sica i Enginyeria Nuclear, EUETIB,
  Universitat Polit\`{e}cnica de Catalunya, C/~Comte d'Urgell 187,
  E-08036 Barcelona, Spain
  \and
  Institut d'Estudis Espacials de Catalunya (IEEC), Ed. Nexus-201,
  C/~Gran Capit\`{a} 2-4, E-08034
  Barcelona, Spain
  \and
  Departament de F\'{\i}sica Aplicada, Universitat Polit\`{e}cnica
  de Catalunya, C/~Esteve Terrades, 5,
  E-08860 Castelldefels, Spain
  \and
  Facultad de Ciencias Astron\'omicas y Geof\'{\i}sicas,
  Universidad Nacional de La Plata, Paseo del Bosque
  s/n, (1900) La Plata, Argentina}

\abstract{The origin of R Coronae Borealis stars has been elusive for
  over 200 years. Currently, two theories for their formation have
  been presented. These are the Final Flash scenario, in which a dying
  asymptotic giant branch (AGB) star throws off its atmosphere to
  reveal the hydrogen poor, heavily processed material underneath, and
  the double degenerate scenario, in which two white dwarfs merge to
  produce a new star with renewed vigour. Some theories predict that
  the temperatures reached during the latter scenario would destroy
  any lithium originally present in the white dwarfs. The observed
  lithium content of some R Coronae Borealis stars, therefore, is
  often interpreted as an indication that the Final Flash scenario
  best describes their formation.  In this paper, it is shown that
  lithium production can, indeed, occur in the merging of a helium
  white dwarf with a carbon-oxygen white dwarf if their chemical
  composition, particularly that of $^{3}$He, is fully considered. The
  production mechanism is described in detail, and the sensitivity of
  lithium production to the merger environment is investigated.
  Nucleosynthesis post-processing calculations of smoothed-particle
  hydrodynamics (SPH) tracer particles are performed to show that any
  lithium produced in these environments will be concentrated towards
  the cloud of material surrounding the R~CrB star. Measurements of
  the lithium content of these clouds would, therefore, provide a
  valuable insight into the formation mechanism of R~CrB stars.}


\maketitle

\section{Introduction}
Some chemically peculiar stars are deficient in hydrogen, while being
enriched in carbon and oxygen. They are known collectively as
hydrogen-deficient stars, and can be broken into three sub-groups:
Hydrogen Deficient Carbon (HdC), Extreme Helium (EHe), and R Coronae
Borealis (R CrB) stars. Their origin is difficult to explain through
standard stellar evolution theory, so alternative scenarios for their
formation are needed.  One such theory is the 
Final Flash (FF) scenario, in which a late helium-shell flash occurs
in a post-Asymptotic Giant Branch (AGB) star, hence moving it back
towards the AGB part of the Hertzsprung-Russell diagram. The numerical
models of \cite{SCH79} first showed that fully developed helium shell
flashes at temperatures greater than 100 MK are possible, lending
weight to this theory. Indeed, FF objects, such as Sakurai's
  object, are known to exist~\citep{DUE96}. The other leading theory
is known as the double degenerate (DD) scenario, which involves the
merging of two white dwarf stars -- one with a carbon-oxygen (CO)
core, the other, helium (He). Full understanding of these two
production mechanisms is yet to be reached, and considerable effort is
still needed to explain all of the observational signatures of these
stars.

A valid question that arises when considering the DD scenario is as
follows: How feasible is it that two white dwarfs with the required
compositions and masses merge to form an object resembling an R~CrB
star? To answer this question, the rate of these events within an
observable distance can first be estimated. Roughly two of every three
stars are born in binary systems~\citep{NEL01}. Of these stars,
however, only a fraction are expected to evolve into double white
dwarf systems that will merge within one Hubble time. Various
estimates for this fraction are available, but as an example, the
models of \cite{NEL01} suggest that the birth rate of double white
dwarfs in our galaxy is around 0.05 yr$^{-1}$, of which around 50\%
will merge within one Hubble time. About 15\% of those merging systems
will comprise of a carbon-oxygen and a helium white dwarf (where the
CO white dwarf was formed first). While direct comparisons of this
rate with the population of known hydrogen deficient stars is not
trivial, these rates are consistent with the known population of close
binary systems~\citep{IBE97}.

The DD scenario is also successful, at least qualitatively, in
explaining the bulk of the surface abundances in hydrogen deficient
stars~\citep[e.g.,][]{JEF11}. Recently, models with detailed nuclear
networks have shown that nucleosynthesis can occur in the DD scenario,
thus explaining the over-abundance of $^{19}$F and $^{18}$O in R~CrB
stars~\citep{LON11}. However, some details such as the enrichment of
s-process elements are more difficult to explain. Lithium abundances
also pose a problem, with four stars analysed by \cite{ASP00} yielding
over-abundances in the range of 1.4 to 2.3 dex. It has been
hypothesised that lithium can be enriched to these levels successfully
in the FF scenario, while being hard to reconcile with high oxygen
abundances in the DD scenario~\citep{ASP00,CLA11}. The authors of
those studies therefore interpreted the success of lithium production
in the FF scenario as evidence that hydrogen deficient stars are most
likely created in the dying stages of AGB stars. However, the
formation of lithium in the DD scenario has not been fully explored to
date.

The aim of this paper is to show that lithium can, indeed, be produced
in the merging of a helium white dwarf with a carbon-oxygen white
dwarf. We will begin in Sect.~\ref{sec:prior-evol} by discussing the
evolution of main sequence stars into a white dwarf binary system
before explaining the mechanism necessary for lithium enrichment in
Sect.~\ref{sec:theory}. A number of numerical tests of this production
are presented in Sect.~\ref{sec:tests} and its sensitivity to the
thermodynamic properties of the merging event is discussed. These
numerical tests will be compared with a detailed nucleosynthesis study
by using Smoothed Particle Hydrodynamics tracer particles in
Sect.~\ref{sec:models} and we will conclude in Sect.~\ref{sec:concl}.

\section{Prior evolution of the binary system}
\label{sec:prior-evol}
In order to understand the mechanism responsible for producing lithium
in the DD scenario, the evolution of the binary system must first be
addressed. The most likely system giving rise to the necessary white
dwarf pair (one $0.6 \, M_{\odot}$ carbon-oxygen and one $0.4 \,
M_{\odot}$ helium white dwarf) is a system containing stars with
masses of $5  \, M_{\odot}$ and $1 \, M_{\odot}$, respectively. Following two common
envelope episodes, mass is ejected from the system resulting in two
white dwarf stars in a close orbit~\citep[see][for more
information]{IBE84,NEL01}.

Before considering the nucleosynthesis that occurs during the merger,
the compositions of the white dwarfs must be understood. It has been
pointed out \citep{SAI02,JEF11} that the surface composition of
merging white dwarfs is different from their core compositions. This
is because during the mass loss phases of the stars, not all unburned
material is lost in stellar winds, thus thin, partially
burned layers remain on the white dwarf surfaces. Therefore,
some buffer regions were defined in those studies to take
this material into account. This material is partially processed
because it was originally close to the hydrogen burning shell in the
progenitor star, so its chemical make-up should be altered from its
original composition.

To estimate the chemical abundances in the partially processed layer,
let us consider how hydrogen burns in these environments. For the
helium white dwarf (originally a $1 \, M_{\odot}$ star), hydrogen should be
consumed through the CNO cycle (dominating the energy production) and
the pp-chain, which converts four protons into helium while releasing
energy in the form of photons and neutrinos. The pp-chain follows
three paths depending on the temperature of the environment. The
dominant proton-burning paths with respect to ascending temperature
are:
\begin{description}
\item[PPI\phantom{II}] p(p,$\beta^+$)d(p,$\gamma$)$^3$He($^3$He,2p)$^4$He,
\item[PPII\phantom{I}] p(p,$\beta^+$)d(p,$\gamma$)$^3$He($^4$He,$\gamma$)$^7$Be($e^-,\nu$)$^7$Li(p,$^4$He)$^4$He,
\item[PPIII] p(p,$\beta^+$)d(p,$\gamma$)$^3$He($^4$He,$\gamma$)$^7$Be(p,$\gamma$)$^8$B($\beta^+$)$^8$Be$\rightarrow$2$^4$He.
\end{description}
A series of partial differential equations can be written to describe
these chains \citep[see, for example,][]{ILIBook}. Of particular
interest to us are those governing the evolution of $^3$He as will
become apparent in Sect.~\ref{sec:theory}.  By assuming that
deuterium quickly reaches its equilibrium abundance in the star, the
equation governing $^3$He evolution can be written as
\begin{equation}
  \label{eq:he3-pde}
  \frac{\textrm{d}(^3\textrm{He})}{\textrm{d}t} =  \displaystyle \frac{(\textrm{H})^2}{2}\langle \sigma v \rangle_{pp} -
  (^3\textrm{He})^2\langle \sigma v \rangle_{33} - (^3\textrm{He})(^4\textrm{He})\langle \sigma v \rangle_{34},
\end{equation}
where $(\textrm{H})$, $(^3\textrm{He})$, and $(^4\textrm{He})$ are the
abundances of hydrogen, $^3$He, and $^4$He, respectively; $\langle
\sigma v \rangle_{pp}$ is the rate of the p(p,$\beta^+$)d reaction;
$\langle \sigma v \rangle_{33}$ is the rate of the
$^3$He($^3$He,2p)$^4$He reaction; and $\langle \sigma v \rangle_{34}$
is the reaction rate of $^3$He($^4$He,$\gamma$)$^7$Be.  The
equilibrium abundance of $^3$He, $(^3\textrm{He})_e$, is determined by
finding the roots of this quadratic equation when
$d(^3\textrm{He})/\textrm{d}t=0$, the only physical solution being
\begin{eqnarray}
  (^3\textrm{He})_e &=& \displaystyle \frac{1}{2
    \langle \sigma v \rangle_{33}}\left[-(^4\textrm{He})\langle \sigma v \rangle_{34} + \right. \nonumber \\
    &&\left. \sqrt{2(\textrm{H})^2\langle \sigma v \rangle_{pp}\langle \sigma v
        \rangle_{33}+(^4\textrm{He})^2 \langle \sigma v \rangle_{34}^2}\,
  \right] \label{eq:3he-equil}
\end{eqnarray}
Hydrogen shell burning in a red giant branch star with approximately
$1 \, M_{\odot}$ occurs at 10-20~MK, thus energy is produced primarily
through the CNO cycle. However, the pp-chain is still active in these
conditions with a minor contribution to energy production and is
essential for the production of $^3$He, and hence $^7$Li. Throughout
the following discussion, a point in the hydrogen burning shell
defined as the position at which $H=0.5$ and $(^4\textrm{He})=0.5$ is
considered\footnote{Note that in reality, the outer surface of the
  helium white dwarf does not consist of a discontinuity of
  abundances, but rather a continuous profile of abundances with
  respect to radius}. This point, with an estimated temperature of $T
= 15$~MK, is destined to become part of the hydrogen buffer in both
white dwarfs once their envelopes have been lost to stellar winds. By
using these values in Eq.~(\ref{eq:3he-equil}), $^3$He mass fractions
of $(^3\textrm{He})_e = 1 \times 10^{-5}$ are found to be reached in
$5 \times 10^5$~years, well below the red giant branch lifetime for
these stars. Note that the bulk of $^3$He is produced in the envelope
of the star at much larger radii (and lower temperatures) than
considered here. At later stages of the star's evolution, the envelope
material is convectively mixed into this region, consequently
\textit{increasing} the $^3$He content relative to what is produced in
the hydrogen-burning shell. In this section, we do not take these
time-dependent effects into account, but rather take the conservative
approximation that no convective mixing takes place to increase $^3$He
in the partially H-burned shell. These approximations are only made in
Sects.~\ref{sec:prior-evol}--\ref{sec:tests} in order to illustrate
the mechanism for producing $^7$Li in white dwarf mergers. In
Sect.~\ref{sec:models}, these approximations are relaxed to take the
full evolutionary history of the white dwarfs into account. In the
centre of the hydrogen burning shell in a $1 \, M_{\odot}$ red giant
star, one would therefore expect $^3$He to be overabundant. Later,
once the star's envelope has been eroded during the common envelope
stage, this $^3$He persists in the surface layers of both
white dwarfs. The fate of this material becomes essential in
understanding lithium production during white dwarf mergers.

\section{Merging event}
\label{sec:theory}
Following the common envelope stages, gravitational wave emission is
responsible for further reducing angular momentum in the system until
finally, once tidal forces become strong enough, the helium white
dwarf is disrupted rapidly (in about 100 seconds) and is accreted on
to the carbon-oxygen white dwarf. This merging episode results in an
object that consists of three parts: (i) a dense, central object
consisting of the core of the carbon-oxygen white dwarf; (ii) a hot
corona in which nucleosynthesis has occurred; and (iii) a thick
accretion disk~\citep[e.g.,][]{LOR10}. While the hot corona should
contain some material originating in the outer regions of the
carbon-oxygen white dwarf, thus explaining the high abundances of
carbon and oxygen in hydrogen deficient stars, it consists primarily
of material from the helium white dwarf. 

As we have shown in Sect.~\ref{sec:prior-evol}, the hydrogen rich
buffer of the white dwarfs is expected to be enriched in $^3$He. This
is of particular interest because helium burning conditions are
reached during the merger, so the unstable nucleus $^7$Be can be
synthesised, which subsequently captures an electron to produce
$^7$Li.  This synthesis depends on rapid heating and cooling of
material on timescales that are rare in astrophysical environments
(see Sect.~\ref{sec:tests}). The production reaction chain itself is
simple: $^3$He($\alpha$,$\gamma$)$^7$Be($e^-,\nu$)$^7$Li. It is
complicated somewhat because $^7$Be can be destroyed by proton capture
to produce $^8$B, which subsequently decays back into two
$\alpha$-particles or undergo further proton capture to produce
$^9$C. Furthermore, at high temperatures, the reverse reaction
$^8$B($\gamma$,p)$^7$Be is particularly important for $^7$Be
production~\citep[e.g.,][]{BOF93,HER96}. Nevertheless, at low
temperatures, the partial differential equations discussed in
Sect.~\ref{sec:prior-evol} are consulted once again to provide us with
the $^7$Be equilibrium abundance:
\begin{equation}
  \label{eq:7Be-equil}
  (^7\textrm{Be})_e \approx
  \frac{(^3\textrm{He})_e(^4\textrm{He})\langle \sigma v \rangle_{34}}{\lambda_{7} + H\langle \sigma v \rangle_{17}}
\end{equation}
where $\lambda_{7}$ is the electron capture rate of $^7$Be into $^7$Li
and $\langle \sigma v \rangle_{17}$ is the rate of the
$^7$Be(p,$\gamma$)$^8$B reaction.

The behaviour of this equilibrium abundance and its timescale are
important in $^7$Li production. As the stellar temperature rises,
so does the ($^7$Be)$_{e}$ abundance, and the time required to
reach those equilibrium values decreases dramatically. At a
temperature of $T=100$~MK, therefore, equilibrium abundances of
$^7$Be$_{e}=1 \times 10^{-7}$ are reached in about one second for
material originating from the centre of the hydrogen
buffer. Therefore, if high temperatures are reached followed by rapid
cooling before exhaustion of the $^3$He reservoir (the
time-scales involved do not allow for replenishment of $^3$He),
$^7$Be will freeze-out at highly enriched values. This
$^7$Be is then free to undergo electron capture, producing
$^7$Li with a half-life of about 53 days~\citep{Nubase},
resulting in a stellar environment enriched in $^7$Li. 

We have shown that, in principle, $^7$Be (and consequently
$^7$Li) can be produced in high concentrations if the environment
reaches the required temperatures for the correct time period. The
obvious question that arises, therefore, is: how sensitive is this
$^7$Be production to the thermodynamic properties of the merging
material?


\section{Numerical tests}
\label{sec:tests}
In the case of a white dwarf merger, the buffer layer of the helium
white dwarf is expected to be amongst of the first material to impact
the surface of the carbon-oxygen white dwarf, thus reaching high
temperatures -- typically above 300~MK~\citep{LOR10}. The energy
generated by nuclear burning at these temperatures ejects the
processed material back out to larger orbits, cooling it and
allowing for any $^7$Be to decay into $^7$Li. The assumed temperature
density profile for buffer material falling on to the surface
of a carbon-oxygen white dwarf therefore consists of three
stages: (i) a rapid exponential rise; (ii) a brief stage in which both
the temperature and density remain constant; and (iii) an exponential,
adiabatic cooling stage.


A small hydrogen burning network is used consisting of nuclei from A=1
to A=12 to study the $^7$Li production sensitivity on the profile
parameters. The network contains all reactions pertinent to
nucleosynthesis in the pp-chain, including reverse reactions.  Rates
are adopted from the REACLIB database \citep{CYB10}. In order to
approximate the production of $^7$Li in the merger, the same location
in the hydrogen buffer as used in Sect.~\ref{sec:theory} is
considered. This region has $^1$H and $^4$He abundances of
$(\textrm{H}) = (^4\textrm{He}) = 0.5$ and a $^3$He abundance of
$(^3\textrm{He}) = 1 \times 10^{-5}$, as calculated in
Sect.~\ref{sec:prior-evol}. All other nuclei have negligible initial
abundances.

\begin{table}
  \caption{Sensitivity study variables}
  \centering
  \begin{tabular}{c|cc}
    \hline \hline
    Variable & Minimum value & Maximum value \\ \hline
    $t_r$ (s) & 0.01  & 20  \\
    $t_p$ (s) & 0.01  & 50  \\ 
    $t_c$ (s) & 0.1   & 50  \\
    $T_{\textrm{max}}$ (MK) & 80  & 800 \\
    \hline \hline
  \end{tabular}
  \tablefoot{Summary of the variable ranges used for the $^7$Li production sensitivity study. $t_r$ refers to the rise time to maximum temperature, $T_{\textrm{max}}$. $t_p$ is the time spent at this peak temperature, and $t_c$ is the time taken to cool back to $T=1$~MK.}
  \label{tab:vars}
\end{table}

\begin{figure}
  \centering
  \includegraphics[width=\hsize]{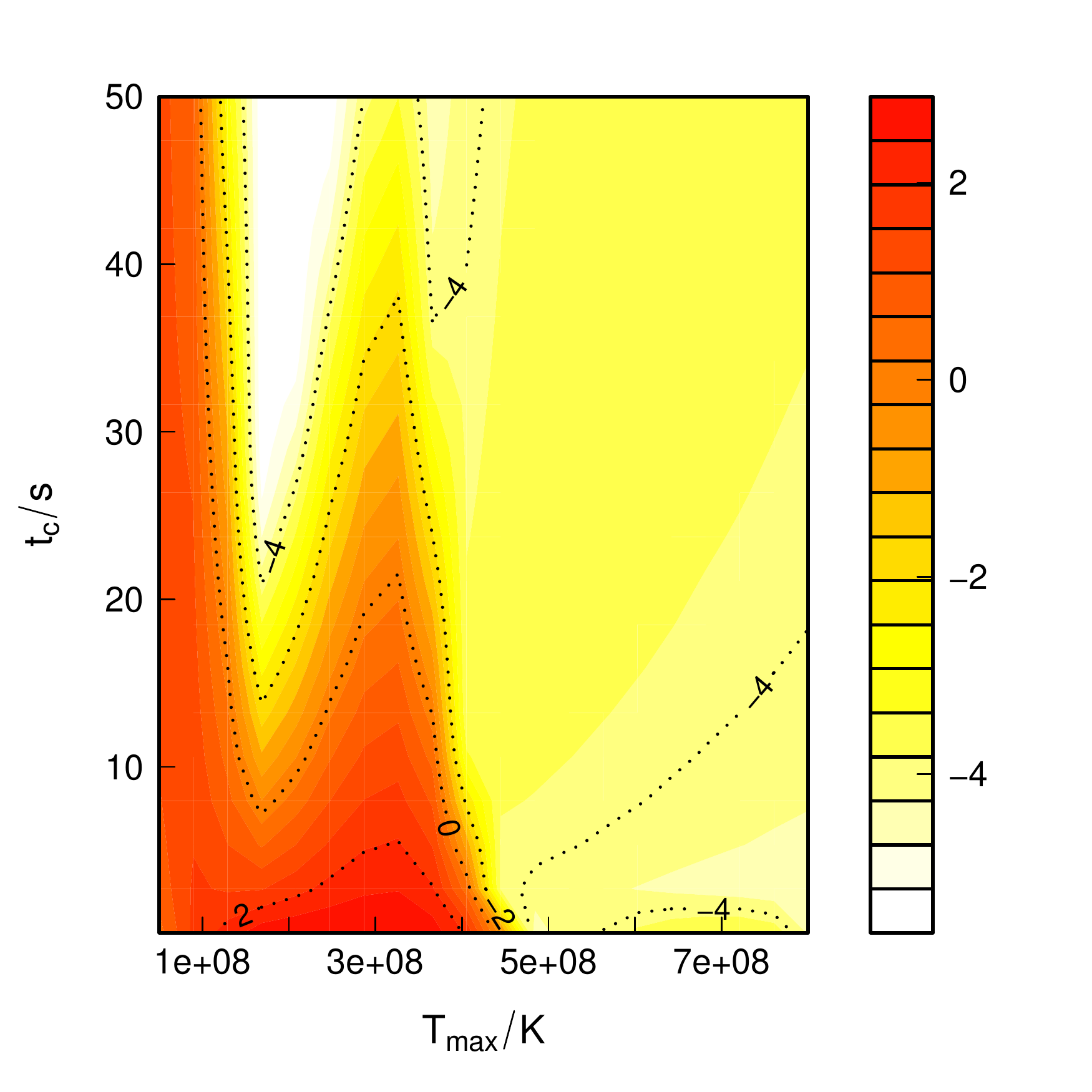}
  \caption{(Colour online) Sensitivity of $^7$Li production to the
    cooling time, $t_c$, and maximum temperature, $T_\textrm{max}$, of
    the profile. The contour labels refer to $\log_{10}
    (^7\textrm{Li})/(^7\textrm{Li})_{\odot}$ (i.e., zero represents a
    lithium abundance equal to solar values). These values have been
    calculated assuming a rise time of $t_r=1$~s and a peak-time of
    $t_p=1$~s.}
  \label{fig:tcVsTmax}
\end{figure}

The $^7$Li production sensitivity is investigated with respect to four
profile parameters: (i) rise time, $t_r$, taken to heat the material
from $T=1$~MK to the maximum temperature, $T_{\textrm{max}}$; (ii)
cooling time, $t_c$ (i.e., the time taken to cool to $T=1$~MK); (iii)
amount of time spent at peak temperature, $t_p$; and (iv) maximum
temperature $T_{\textrm{max}}$. $^7$Li production is also
sensitive to the initial abundances, depending mostly on the initial
concentration of $^3$He. The parameter ranges adopted are summarised
in Table~\ref{tab:vars}. The density is chosen to vary from
$\rho_0=1 \times 10^3$~g/cm$^{3}$ to $\rho_{\textrm{max}}=1 \times
10^5$~g/cm$^3$, with a cooling tail that follows an adiabatic cooling
curve. This behaviour is consistent with the SPH models of
\cite{LOR10}. An equally spaced grid of 20 values are used
for each variable, thus nucleosynthesis arising from 160\,000 profiles
is analysed. A maximum $^7$Li abundance of $(^7\textrm{Li}) =
1.4 \times 10^{-5}$ (corresponding to a factor of 1000 more lithium
than is found in our solar system) occurs at temperatures of 350~MK
with the fastest rise, cooling, and peak times considered. These
extreme values are, however, unrealistic. In reality, rapid heating of
the material is expected, but the cooling time is subject to
variability depending on the dynamical situation of the merger. We
therefore consider the dependence of $^7$Li production to peak
temperature and cooling time assuming a nominal rise and peak times of
1~s. This sensitivity is shown in Fig.~\ref{fig:tcVsTmax}. A number of
interesting features are apparent. First, lithium is produced
most efficiently for sharp profiles (i.e., short rise and cooling
times). This is because of the freeze-out mechanism discussed in
Sect.~\ref{sec:theory} coupled to the fact that $^3$He replenishment
cannot occur quickly enough to maintain the $^3$He:$^7$Be equilibrium
at peak
temperatures. 
Secondly, the production of lithium as a function of temperature seems
to follow a rather complicated pattern that requires some detailed
discussion.

By considering Fig.~\ref{fig:tcVsTmax}, three temperature regions are
defined according to how much $^7$Li is produced: (i) the low
temperature region in which $^7$Li production is independent of
fall time; (ii) the medium temperature region of $200 <
T_{\textrm{max}} < 400$~MK, in which the $^7$Li abundance depends
strongly on the profile shape and peaks around
$T_{\textrm{max}}=300$~MK; and (iii) the high temperature region in
which very little $^7$Li is produced.

In the low temperature regime, around $T_{\textrm{max}}=100$~MK, the
rate of the $^3$He($\alpha$,$\gamma$)$^7$Be reaction is not high
enough to fully destroy $^3$He during the merger. Therefore, an
equilibrium is reached, which is rather insensitive to the dynamics
affecting the shape of the profile. The equilibrium $^7$Be abundance
then follows the behaviour discussed in
Sect.~\ref{sec:theory}. However, as the temperature rises, the
reaction rates increase, thus causing $^3$He to be partially,
and eventually fully destroyed. In this event, $^7$Be can only be
replenished by the photo-disintegration reaction,
$^8$B($\gamma$,p)$^7$Be, whose rate is low around 200~MK, thus
explaining the rapid drop in $^7$Li production from 150~MK to 200~MK.

In the medium temperature regime, $^7$Be can no longer be produced
from $^3$He, and the equilibrium ratio of $^7$Be/$^8$B is
approximately proportional to the ratio of their reaction rates:
$\lambda_{87} / \langle \sigma v \rangle_{78}$ (where $\lambda_{87}$
is the photo-disintegration rate of $^8$B into $^7$Be and $\langle
\sigma v \rangle_{78}$ is the proton capture rate of
$^7$Be(p,$\gamma$)$^8$B). Assuming that all of the $^3$He is
converted to $^7$Be and $^8$B early in the merging event, and
noticing that the reaction rate ratio increases with temperature, it
follows that the surviving $^7$Be should increase with
temperature. However, one further effect needs addressing: the
destruction of $^8$B by the $^8$B(p,$\gamma$)$^9$C reaction. This
reaction, active above $T=100$~MK, erodes the $^8$B abundance and thus
the abundance of $^7$Be. For this reason, longer profiles
(i.e., profiles that take longer to cool) suffer from lower $^7$Li
production.

At high temperatures, above $T=400$~MK, another process comes into
play: the destruction of $^7$Be by the
$^7$Be($\alpha$,$\gamma$)$^{11}$C reaction. This reaction
serves to divert the reaction flow away from $^8$B, and thus, the
equilibrium $^7$Be/$^8$B cannot be established. The rapid
increase of the $^7$Be($\alpha$,$\gamma$)$^{11}$C rate as a
function of temperature causes the sharp cut-off in $^7$Li
production shown in Fig.~\ref{fig:tcVsTmax}. Therefore, in helium rich
environments, $^7$Be is not expected to survive destruction by
this mechanism at temperatures above $T=400$~MK.

\section{Results from hydrodynamic models}
\label{sec:models}
\subsection{White dwarf models}
\label{sec:wd-models}
As shown in the preceding sections, the initial composition of the
white dwarfs is critical to understanding lithium production during
the merger event. Therefore, the detailed white dwarf models of
\cite{ALT10} and \cite{REN10} are used to evolve stars from the main
sequence to the white dwarf stage. These evolutionary models contain a
limited nuclear network of 16 elements: $^1$H, $^2$H,
$^3$He, $^4$He, $^7$Be, $^{12}$C, $^{13}$C,
$^{14}$N, $^{15}$N, $^{16}$O, $^{17}$O, $^{18}$O,
$^{19}$F, $^{20}$Ne, and $^{22}$Ne. This network is
sufficient for following nuclear energy generation during the
evolution of the star.

The carbon-oxygen white dwarf model is obtained by evolving
$2.5 \, M_{\odot}$ main sequence stars with metallicities of $Z=1.5 \times
10^{-2}$ and $Z = 1 \times 10^{-5}$ through the thermally pulsing AGB
stage and subsequent mass loss period \citep[using the prescription
of][]{VAS93} through the planetary nebula phase to the white dwarf
phase. The subsequent cooling of the white dwarf is carefully modelled
\citep[see][for more information]{ALT10} to take into account the
effects of, among others, diffusion of material. 

The helium white dwarf model is computed similarly to the
carbon-oxygen white dwarf. However, since the orbital radius of the
two stars decreases during the first common envelope stage (i.e., when
the more massive star loses mass in its AGB phase), the common
envelope could occur earlier in the stars evolutionary path. A $2.5 \,
M_{\odot}$ main sequence star is therefore used. During the stars
evolution on the red giant branch, an artificially high mass-loss rate
is applied to account for the common envelope stage. As a consequence
of this high mass-loss rate, the star loses its envelope
before core helium burning commences, and evolves on to the
helium white dwarf cooling track.

Following the evolution described above, white dwarf models are
obtained that describe the distribution of isotopes in the two stars
as a function of radius. These models are then mapped onto our
hydrodynamic models in order to follow the evolution of these
abundances as the two white dwarfs undergo a merging event.

\subsection{Merging system models}
\label{sec:merger}
The hydrodynamic models used to provide us with input to our
nucleosynthesis calculations are the same as those presented in
\cite{GUE04}, \cite{LOR09,LOR10}, and\ \cite{LON11}. A Lagrangian
Smoothed Particle Hydrodynamics (SPH) code is used to obtain
temperature-density profile input for the post-processing nuclear
network. The model used for the calculations presented here
is computed for the merging of a $0.4 \, M_{\odot}$ helium
white dwarf with a carbon-oxygen white dwarf of mass $0.8 \,
M_{\odot}$. While this model does not correspond exactly with the
white dwarf masses required to form an R CrB star, their dynamical
merging nature should be similar. Additionally, since the envelope of
the final system is mostly comprised of material originating in the
helium white dwarf, this merging system should reproduce sufficiently
the $0.4+0.6 \, M_{\odot}$ case. The SPH particle masses were $2.6
\times 10^{-6} \, M_{\odot}$ and $5.3 \times 10^{-6} \, M_{\odot}$ for
the two stars, respectively to provide a total of 300\,000 particles.

A sample of 10\,000 tracer particles is chosen to follow the chemical
evolution of material during the merger. Following the merging event,
only material that is present in the hot corona and surrounding
accretion disk will be visible to any observer. Consequently, only
particles that finish their evolution with orbital radii outside the
dense central object are chosen as tracers. These particles correspond
to those that would form the atmosphere and chromosphere\
\citep{CLA96} of a hydrogen deficient star. A similar nuclear network
as that presented in Sect.~\ref{sec:tests} (with additional isotopes
introduced to account for those in our white dwarf models) is used,
which is sufficient to model the synthesis of light
elements. Elemental abundances, for comparison with observational
data, are then calculated by summing the isotopic abundances obtained
from the models.

\subsection{Model results}
\label{sec:mod-results}
\begin{figure}
  \centering
  \includegraphics[width=0.8\hsize]{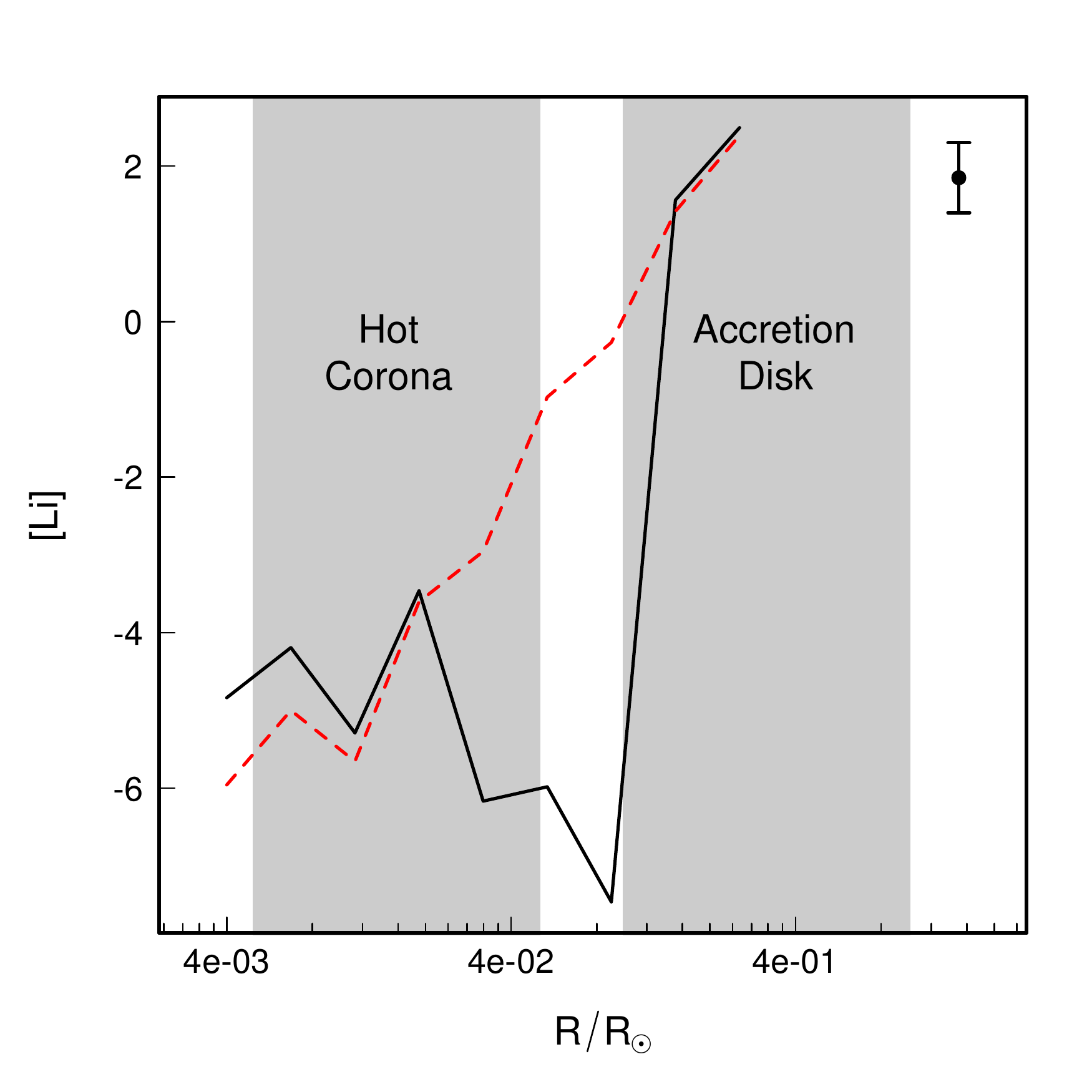}
  \caption{(Colour online) Lithium abundance yield obtained from our
    SPH models as a function of radius in the post-merger object. The
    abundances are represented as logarithmic abundances relative to
    solar values i.e., $[\textrm{Li}] =
    \log_{10}((\textrm{Li})/(\textrm{Li})_{\odot})$ and
    $(\textrm{Li})$ is the mass fraction of lithium. $R/R_{\odot}$
    represents the radius in units of solar radii. The solid black
    line represents the radial lithium abundance resulting from the
    merging of solar metallicity stars, while the dashed red line
    represents that of $Z=1\times 10^{-5}$ stars. The single point
    with error bars represents the range of observed lithium
    abundances in R~CrB stars~\citep[see][and references
    therein]{JEF11}.}
  \label{fig:mod-results}
\end{figure}

Average lithium abundances, normalised to solar values, as a function
of radius in the merger product are shown in
Fig.~\ref{fig:mod-results}. Clearly, lithium enrichment is present in
the outer regions of the object with respect to solar abundances. The
radii at which this high enrichment are found are, indeed, rather
large, and correspond most likely to the accretion disk surrounding
the merger product. The reason for this enrichment is that the buffer
material, which is destined to produce lithium because of its high
concentration of $^3$He, is amongst the first matter accreted
on to the carbon-oxygen white dwarf. Consequently, it will
heat up rapidly and is forced out to larger orbital
radii. Subsequent accretion of helium rich matter from the destroyed
helium white dwarf prevents this material from falling back
on to the carbon-oxygen white dwarf's surface and it
remains in the accretion disk following the merging event.

Lithium production is also expected to be independent of metallicity
(it depends mainly on the $^3$He enrichment of the white
dwarfs). Indeed, this is the case in the low metallicity models
considered (represented by a dashed red line in
Fig.~\ref{fig:mod-results}) where the lithium abundance found in the
outer regions of the final object agree very well with those found
with our high metallicity models.


The distribution of lithium in our models could explain why lithium is
only observed in a limited number of R CrB stars. Although no
accretion disks have been detected, our SPH models suggest that a
rapidly rotating Keplerian disk should be formed subsequent to the
merging event~\citep[see][and references therein]{LOR10}. This disk
could be partially responsible for the observed circumstellar material
surrounding R CrB stars. If this is the case, then if an observer were
to view the object side-on, the accretion disk could obscure the
atmosphere of the star and dominate any spectra obtained. If the star
were face-on to the observer, however, radiation from the surface of
the star itself will dominate the observers measurements, and hence,
lithium abundance measurements would be rather low. Our discovery for
the distribution of lithium in the merger product could also be used
observationally to distinguish between the FF and DD merger
scenarios. If, indeed, the DD scenario was the formation channel for
the star, our models indicate that lithium will be present primarily
in the cloud of material surrounding the star. The leading theory
explaining the characteristic declines of these stars~\citep{CLA96}
suggests that the dramatic drop in luminosity of the star is caused by
a ``puff'' of carbon-rich material that is being accelerated towards
the observer. If the star can be observed early during one of these
declines, the surface of the star will be partially eclipsed, a
process known as ``veiling''~\citep{LAM90}, hence revealing the
diffuse surrounding material for measurement. If lithium is present in
high quantities in this material, it could indicate that the DD
scenario is, indeed, responsible for the formation of R CrB stars.

\section{Conclusions}
\label{sec:concl}
In this paper, we have investigated the production of lithium in hot
mergers of helium and carbon-oxygen white dwarf stars. Both by
performing analytical calculations and by post-processing
nucleosynthesis of SPH tracer particles, significant lithium
production is possible.

In our analytical models, we have shown that this production is
strongly dependent on the dynamical properties of the material in the
merger and have placed constraints on the conditions required for
lithium production. To produce large amounts of lithium the material,
which was enriched in $^3$He in prior evolution, must reach high
temperatures rapidly, and then cool rather rapidly also. Lithium
production also depends sensitively on the maximum temperature reached
by the material through a rather complicated interplay of reactions
and their reverse counterparts. Maximum lithium yields are achieved at
peak temperatures of around $T_{\textrm{max}}=300$~MK provided the
material undergoes rapid heating and cooling episodes.

Post-processing of tracer particles from SPH models of merging white
dwarfs confirms that lithium can, indeed, be produced in the DD
scenario, thus satisfying the constraints investigated in our
analytical models. By performing these calculations for the merging of
a $0.4 \, M_{\odot}$ helium white dwarf with a $0.8 \, M_{\odot}$
carbon-oxygen white dwarf, we have shown that the agreement between
our models and the observed lithium surface abundance in hydrogen
deficient stars is reassuring. Furthermore, our models predict that
lithium should be predominantly produced in the outer regions of the
merger product. If measurements are made to identify the location of
lithium in R~CrB stars, they would provide a vital clue in determining
the formation mechanism of these peculiar stars. Further studies
should also be performed to investigate the influence of other factors
such as the white dwarf masses, compositions, and dynamical properties
on the nucleosynthesis of lithium in white dwarf mergers.

It was previously believed that lithium observations in some R CrB
stars indicated that they are produced by a powerful final helium
flash in a dying AGB star. As we have shown in this paper, lithium can
be produced in the merging of two white dwarfs if the conditions meet
the criteria laid out here. Lithium cannot, therefore, be used as a
strong indicator that the DD formation scenario is
invalid for R~CrB stars.

\begin{acknowledgements}
  This work has been partially supported by the Spanish grants
  AYA2010-15685 and AYA2011-23102, by AGAUR grant SGR1002/2009, by the
  E.U. FEDER funds, and by the ESF EUROCORES Program EuroGENESIS
  through the MICINN grant EUI2009-04167.
\end{acknowledgements}

\bibliographystyle{aa}
\bibliography{bib}

\end{document}